\documentclass[journal]{IEEEtran}

\usepackage{cite,graphicx,subfigure,hyperref,amsmath,amsthm,amsfonts,array}

\newlength{\Lpr}
\newsavebox{\Bpr}

\newcommand{\V}[1]{\mbox{\boldmath$\mathbf{#1}$\unboldmath}}

\newcommand{\sinc}{\ensuremath{{\mathrm{sinc}}}}

\newcommand{\bdm}{\begin{displaymath}}
\newcommand{\edm}{\end{displaymath}}

\newcommand{\be}[1]{\begin{equation} \label{#1}}
\newcommand{\ee}{\end{equation}}

\newcommand{\bae}[3]{
\begin{equation} \label{#1}
\renewcommand{\arraystretch}{#2}
\begin{array}{#3}}

\newcommand{\eae}{\end{array}\end{equation}}

\newcommand{\baen}[2]{
\begin{displaymath} 
\renewcommand{\arraystretch}{#1}
\begin{array}{#2}}

\newcommand{\eaen}{\end{array}\end{displaymath}}

\newcommand{\DefLetter}[4]{
\newcommand{#1}{\ensuremath{\V{#2}}} 
\newcommand{#3}{\ensuremath{\V{#4}}} 
}

\DefLetter{\vzer}{0}{\mzer}{0}
\DefLetter{\vone}{1}{\mone}{1}
\DefLetter{\va}{a}{\ma}{A}
\DefLetter{\vb}{b}{\mb}{B}
\DefLetter{\vc}{c}{\mc}{C}
\DefLetter{\vd}{d}{\md}{D}
\DefLetter{\ve}{e}{\me}{E}
\DefLetter{\vf}{f}{\mf}{F}
\DefLetter{\vg}{g}{\mg}{G}
\DefLetter{\vh}{h}{\mh}{H}
\DefLetter{\vi}{i}{\mi}{I}
\DefLetter{\vj}{j}{\mj}{J}
\DefLetter{\vk}{k}{\mk}{K}
\DefLetter{\vl}{l}{\ml}{L}
\DefLetter{\vm}{m}{\mm}{M}
\DefLetter{\vn}{n}{\mn}{N}
\DefLetter{\vpr}{p}{\mpr}{P}
\DefLetter{\vq}{q}{\mq}{Q}
\DefLetter{\vr}{r}{\mr}{R}
\DefLetter{\vs}{s}{\ms}{S}
\DefLetter{\vt}{t}{\mt}{T}
\DefLetter{\vur}{u}{\mur}{U}
\DefLetter{\vv}{v}{\mv}{V}
\DefLetter{\vw}{w}{\mw}{W}
\DefLetter{\vx}{x}{\mx}{X}
\DefLetter{\vy}{y}{\my}{Y}
\DefLetter{\vz}{z}{\mz}{Z}

\DefLetter{\vdel}{\delta}{\mdel}{\Delta}
\DefLetter{\vphi}{\phi}{\mphi}{\Phi}
\DefLetter{\vpsi}{\psi}{\mpsi}{\Psi}
\DefLetter{\vrho}{\rho}{\mrho}{\Lambda}
\DefLetter{\vxi}{\xi}{\mxi}{\Xi}

\DefLetter{\valpha}{\alpha}{\malpha}{\Alpha}
\DefLetter{\vbeta}{\beta}{\mbeta}{\Beta}
\DefLetter{\vlam}{\lambda}{\mlam}{\Lambda}
\DefLetter{\vsig}{\sigma}{\msig}{\Sigma}
\DefLetter{\vtau}{\tau}{\mtau}{\tau}
\DefLetter{\vtheta}{\theta}{\mtheta}{\Theta}
\DefLetter{\vome}{\omega}{\mome}{\Omega}
\DefLetter{\vzero}{0}{\mzero}{0}
\DefLetter{\vgam}{\gamma}{\mgam}{\Gamma}
\DefLetter{\veps}{\epsilon}{\meps}{\Epsilon}
\DefLetter{\veta}{\eta}{\meta}{\Eta}

\newcommand{\DefFuncLetter}[2]{
\newcommand{#1}{\ensuremath{{\mathrm{#2}}}} 
}

\DefFuncLetter{\Fzer}{0}
\DefFuncLetter{\Fa}{a}
\DefFuncLetter{\FA}{A}
\DefFuncLetter{\Fb}{b}
\DefFuncLetter{\Fc}{c}
\DefFuncLetter{\FC}{C}
\DefFuncLetter{\Fd}{d}
\DefFuncLetter{\Fe}{e}
\DefFuncLetter{\Ff}{f}
\DefFuncLetter{\Fg}{g}
\DefFuncLetter{\FG}{G}
\DefFuncLetter{\Fh}{h}
\DefFuncLetter{\FH}{H}
\DefFuncLetter{\Fi}{i}
\DefFuncLetter{\Fk}{k}
\DefFuncLetter{\Fl}{l}
\DefFuncLetter{\FL}{L}
\DefFuncLetter{\Fm}{m}
\DefFuncLetter{\Fn}{n}
\DefFuncLetter{\Fnr}{n}
\DefFuncLetter{\FN}{N}
\DefFuncLetter{\Fo}{o}
\DefFuncLetter{\FO}{O}
\DefFuncLetter{\Fpr}{p}
\DefFuncLetter{\FPr}{P}
\DefFuncLetter{\Fq}{q}
\DefFuncLetter{\Fr}{r}
\DefFuncLetter{\Fs}{s}
\DefFuncLetter{\FS}{S}
\DefFuncLetter{\Ft}{t}
\DefFuncLetter{\FT}{T}
\DefFuncLetter{\Fu}{u}
\DefFuncLetter{\FU}{U}
\DefFuncLetter{\Fv}{v}
\DefFuncLetter{\Fw}{w}
\DefFuncLetter{\FW}{W}
\DefFuncLetter{\Fx}{x}
\DefFuncLetter{\Fy}{y}
\DefFuncLetter{\FY}{Y}
\DefFuncLetter{\Fz}{z}
\DefFuncLetter{\FZ}{Z}
\DefFuncLetter{\Falp}{\alpha}
\DefFuncLetter{\Fbet}{\beta}
\DefFuncLetter{\Fchi}{\chi}
\DefFuncLetter{\Fdel}{\delta}
\DefFuncLetter{\Fzet}{\zeta}
\DefFuncLetter{\FEps}{\Epsilon}
\DefFuncLetter{\Feta}{\eta}
\DefFuncLetter{\Fphi}{\phi}
\DefFuncLetter{\FPhi}{\Phi}
\DefFuncLetter{\Fpsi}{\psi}
\DefFuncLetter{\FPsi}{\Psi}
\DefFuncLetter{\Fgam}{\gamma}
\DefFuncLetter{\FGam}{\Gamma}
\DefFuncLetter{\Flam}{\lambda}
\DefFuncLetter{\FLam}{\Lambda}
\DefFuncLetter{\Fsig}{\sigma}
\DefFuncLetter{\Ftau}{\tau}
\DefFuncLetter{\Fome}{\omega}
\DefFuncLetter{\Feps}{\epsilon}
\DefFuncLetter{\Fthe}{\theta}
\DefFuncLetter{\Fvar}{\vartheta}

\DefFuncLetter{\FB}{B}
\DefFuncLetter{\FD}{D}
\DefFuncLetter{\FE}{E}
\DefFuncLetter{\FF}{F}
\DefFuncLetter{\FI}{I}
\DefFuncLetter{\FJ}{J}
\DefFuncLetter{\FM}{M}
\DefFuncLetter{\FR}{R}
\DefFuncLetter{\FV}{V}
\DefFuncLetter{\FX}{X}

\newcommand{\DefCalLetter}[2]{
\newcommand{#1}{\ensuremath{\mathcal{#2}}} 
}
\DefCalLetter{\CC}{C}
\DefCalLetter{\CD}{D}

\DefCalLetter{\CS}{S}
\DefCalLetter{\CV}{V}

\newcommand{\DefSubLetter}[2]{
\newcommand{#1}{\mathrm{#2}} 
}

\DefSubLetter{\slzer}{0}
\DefSubLetter{\sla}{a}
\DefSubLetter{\slA}{A}
\DefSubLetter{\slb}{b}
\DefSubLetter{\slB}{B}
\DefSubLetter{\slc}{c}
\DefSubLetter{\slC}{C}
\DefSubLetter{\sld}{d}
\DefSubLetter{\slD}{D}
\DefSubLetter{\sle}{e}
\DefSubLetter{\slE}{E}
\DefSubLetter{\slf}{f}
\DefSubLetter{\slF}{F}
\DefSubLetter{\slg}{g}
\DefSubLetter{\slG}{G}
\DefSubLetter{\slh}{h}
\DefSubLetter{\slH}{H}
\DefSubLetter{\sli}{i}
\DefSubLetter{\slI}{I}
\DefSubLetter{\slk}{k}
\DefSubLetter{\sll}{l}
\DefSubLetter{\slL}{L}
\DefSubLetter{\slm}{m}
\DefSubLetter{\slM}{M}
\DefSubLetter{\sln}{n}
\DefSubLetter{\slnr}{n}
\DefSubLetter{\slN}{N}
\DefSubLetter{\slo}{o}
\DefSubLetter{\slp}{p}
\DefSubLetter{\slP}{P}
\DefSubLetter{\slq}{q}
\DefSubLetter{\slQ}{Q}
\DefSubLetter{\slr}{r}
\DefSubLetter{\slR}{R}
\DefSubLetter{\sls}{s}
\DefSubLetter{\slS}{S}
\DefSubLetter{\slt}{t}
\DefSubLetter{\slT}{T}
\DefSubLetter{\slu}{u}
\DefSubLetter{\slU}{U}
\DefSubLetter{\slv}{v}
\DefSubLetter{\slw}{w}
\DefSubLetter{\slW}{W}
\DefSubLetter{\slx}{x}
\DefSubLetter{\slX}{X}
\DefSubLetter{\sly}{y}
\DefSubLetter{\slY}{Y}
\DefSubLetter{\slz}{z}
\DefSubLetter{\slZ}{Z}

\DefSubLetter{\slalp}{\alpha}
\DefSubLetter{\slbet}{\beta}
\DefSubLetter{\sldel}{\delta}
\DefSubLetter{\slDel}{\Delta}
\DefSubLetter{\sleps}{\epsilon}
\DefSubLetter{\slgam}{\gamma}
\DefSubLetter{\slphi}{\phi}
\DefSubLetter{\sltau}{\tau}
\DefSubLetter{\slxi}{\xi}
\DefSubLetter{\slthe}{\theta}

\def \sump{\sum_{p=-\infty}^\infty}
\def \summ{\sum_{m=1}^M}
\def \sumk{\sum_{k=0}^{K-1}}
\def \sumq{\sum_{q=q_1}^{q_2}}
\newtheorem*{eprob}{{Estimation problem}}

\begin{document}

\title{Signal Estimation from Nonuniform Samples with RMS Error Bound -- Application to
  OFDM Channel Estimation}

\author{J. Selva   
}

\maketitle

\markboth{}{}

\begin{abstract}

We present a channel spectral estimator for OFDM signals containing pilot carriers,
assuming a known delay spread or a bound on this parameter. The estimator is based on
modeling the channel's spectrum as a band-limited function, instead of as the discrete
Fourier transform of a tapped delay line (TDL). Its main advantage is its immunity to the
truncation mismatch in usual TDL models (Gibbs phenomenon). In order to assess the
estimator, we compare it with the well-known TDL maximum likelihood (ML) estimator in
terms of root-mean-square (RMS) error. The main result is that the proposed estimator
improves on the ML estimator significantly, whenever the average spectral sampling rate is
above the channel's delay spread. The improvement increases with the spectral oversampling
ratio.

\end{abstract}

\section{Introduction}

A basic task in Orthogonal Frequency Division Multiplexing (OFDM) is the estimation of the
channel's spectrum.  The so-called pilot-aided channel estimation (PACE) is an efficient
method for  this task, which consists in first sampling the channel's spectrum
using several pilot carriers, and then interpolating it at the data carrier frequencies
\cite{Hwang09}. There is a large number of references that discuss this kind of spectral
estimation from various points of view, and several surveys
\cite{Tong04,Ozdemir07,Hwang09}.
Just to cite some of the relevant approaches, \cite{Morelli01} analyzes the ML and the
minimum mean-square error (MMSE) estimators. In \cite{Yang01}, the authors apply the
ESPRIT algorithm to this problem assuming a parametric channel model, and in
\cite{Edfors98} the estimation method is based on the Singular Value Decomposition. In
\cite{Xiong13} and \cite{Seo10}, the estimation performance is improved by reducing the
leakage in the truncation of the channel's response. In \cite{Fertl10}, the design of
nonuniform pilot distributions is studied. Finally, there exist letters dedicated to
specific systems like \cite{Yu12b} for DVB-T2.

In PACE, the tapped delay-line (TDL) is the basic analytical tool that allows one to
reduce the estimation problem to that of determining the so-called tap weights. However,
it has the drawback that the discrete channel's response must be truncated at some
indices, thus introducing a mismatch. This truncation is actually an instance of the
well-known Gibbs phenomenon \cite{Gottlieb97}.  In order to overcome this drawback, we
propose in this letter to model the channel's spectrum in PACE in an alternative
way. Instead of using the TDL model, we propose to first expand the spectrum in a sinc
series, and then proceed to minimize the expected root-mean-square (RMS) error assuming a
linear estimator.  The sinc series in this letter requires knowledge about the channel's
delay spread, a parameter that in practice can be either estimated
\cite{Athaudage04} or at least upper bounded from basic considerations about the
propagation channel.

The letter has been organized as follows. In Sec. \ref{sec:bce}, we  analyze the basic
spectral estimation method in OFDM systems that employ pilot carriers, and present the
rationale of the letter. Then, in Sec. \ref{sec:dse} we   derive an estimator for a
generic band-limited signal from nonuniform samples, in which the performance measure is
the RMS error. This estimator will be directly applicable to the problem already discussed
in Sec. \ref{sec:bce} through a proper normalization of the channel's spectrum. Finally,
we will present a numerical example in Sec. \ref{sec:ne} from a well-known reference, in
which a channel spectrum is estimated from pilot carriers in an OFDM system.  In this
example, we will compare the performance of the proposed estimator with that of the ML
estimator based on the usual TDL model.

\subsection{Notation}

In this letter, we will employ the following notation:

\begin{itemize}

\item New symbols or functions will be introduced using ``$\equiv$''.

\item Vectors and matrices will be denoted in lower- and upper-case bold font, respectively,
  ($\vm$, $\mm$).

\item $\mi$ will stand for an identity matrix of proper size. 
 
\item For a given matrix $\ma$ or vector $\va$, $[\ma]_{p,q}$ and $[\vpr]_{r}$
  will respectively denote the $p,q$ component of $\ma$, and the $r$th component of $\va$.

\item $\ma^H$, $\ma^T$ will respectively denote the hermitian and transpose of $\ma$.

\item $\FE\{\cdot\}$ will denote the expectation operator.
\end{itemize}

\section{Basic channel estimation in OFDM from pilot carriers}
\label{sec:bce}

Consider a static channel with impulse response $\Fh(t)$ of finite duration, whose support
is contained in the range $]0,T_h[$. We may view $T_h$ as the channel's delay spread or as
an upper bound on this parameter. If the channel's input is an OFDM signal containing
pilot carriers, the problem of estimating the channel's spectrum can be posed entirely
in the frequency domain, once the usual DFT processing has been performed
\cite[Sec. II]{Hwang09}. Basically, after this processing we may assume that $M$ noisy
samples $V_m$ of the channel's spectrum $\FH(f)$ are available at distinct frequencies
$f_m$ (pilot frequencies). We take this spectrum $\FH(f)$ as the actual spectrum of
the channel, and not as an effective response formed by the channel and the transmit
and receive filters. The samples $V_m$ follow the model
\be{eq:50}
V_m=\FH(f_m)+E_m,
\ee
where $E_m$ are independent complex Gaussian noise samples of variance
$\sigma_E^2$. In general terms, the design of a linear estimator in PACE consists in
determining a set of coefficients $\Fg_m(f)$ such that
\be{eq:51}
\FH(f)\approx \sum_{m=1}^M\Fg_m(f)V_m,
\ee
where the error measure is the expected RMS error given by
\be{eq:52}
\Big(\FE\big\{\big|\FH(f)-\summ V_m \Fg_m(f)\big|^2\big\}\Big)^{1/2}.
\ee

For obtaining $\Fg_m(f)$, the usual approach in the literature consists in approximating
$\FH(f)$ using a TDL model. Specifically, if we truncate an effective discrete response of
the channel, we obtain the approximation
\be{eq:53}  
\FH(f)\approx T\sumq h_{e,q}\Fe^{-j2\pi q T f}, 
\ee
where $T$ is the TDL spacing, $h_{e,q}$ are samples of the discrete response, and $q_1$
and $q_2$ are proper truncation indices.  By substituting this formula into (\ref{eq:51}),
we obtain a signal model with a finite number of unknown parameters $h_{e,q}$,
\be{eq:57}
V_m=T\sumq h_{e,q}\Fe^{-j2\pi q T f}+E_m.
\ee
Finally, if $h_{e,q}$ is identifiable from $V_m$, i.e. if $q_2-q_1+1\leq M$, then we may
approximate $h_{e,q}$ using well-known estimators, like the ML or MMSE estimators and,
finally, interpolate $\FH(f)$ using (\ref{eq:53}), \cite[Sec. III]{Morelli01}. This is the
usual estimation approach in PACE.

Consider now the formula in (\ref{eq:53}). Its  right-hand side
is a truncated Fourier series, which is a suitable tool for approximating 
periodic functions. However, its left-hand side, $\FH(f)$, is hardly ever periodic in
practice. We can see this point by inspecting a typical channel response like
\be{eq:55}
\Fh(t)=\sumk a_k \delta(t-\tau_k),
\ee
for amplitudes $a_k$ and delays $\tau_k$. Its spectrum
\be{eq:56}
\FH(f)=\sumk a_k\Fe^{-j2\pi \tau_k f}
\ee
is not $(1/T)$-periodic, unless all the delays $\tau_k$ are integer multiples of
$T$, a highly unlikely event in practice. The mismatch between the left- and
right-hand sides of (\ref{eq:53}) is no other thing than the well-known Gibbs phenomenon
\cite{Gottlieb97}. For channel modeling, this phenomenon is not relevant, given that we
may always increase the number of taps in (\ref{eq:53}), so confining the Gibbs phenomenon
to small bands close to the frequencies $\pm 1/(2T)$. However, for estimating the
channel's spectrum using (\ref{eq:53}), we have that we cannot increase the number of taps
without limits, because it must be $q_2-q_1+1\leq M$ for the coefficients $h_{e,q}$ to be
identifiable in (\ref{eq:57}). So, if we use the TDL interpolator in (\ref{eq:53}) to
reduce the initial model in (\ref{eq:50}) to that in (\ref{eq:57}), we have introduced a
mismatch. As a consequence, we may expect that statistically efficient estimators for
(\ref{eq:57}), like the ML estimator, do have an additional RMS error component due to
this Gibbs phenomenon we have just described.

In order to eliminate this Gibbs phenomenon, we propose in this letter to replace the TDL
interpolator in (\ref{eq:53}) with a description that better suits the properties of
channel spectra. In simple terms, we propose to model $\FH(f)$ as a band-limited function,
and describe it using a sinc series. More precisely, since the spectrum of $\FH(f)$ is
contained in $]-T_h,0[$  we have that the following series is valid:
\be{eq:54}
\FH(f) =\Fe^{-j\pi f T_h}\sump \FH\Big(\frac{p}{T_h}\Big)(-1)^p\sinc(fT_h-p).
\ee
In contrast with the TDL interpolator in (\ref{eq:53}), this series for $\FH(f)$ is exact,
i.e, there are no truncation errors.  There is, however, a technical nuisance that must be
taken into account when interpreting (\ref{eq:54}). The time content of $\Fh(t)$ must lie
in $]0,T_h[$ and not in $[0,T_h]$, because $\FH(f)$ must be modeled as a bounded function
and not as a finite-energy one; (see \cite[Sec. 6.8]{Higgins96} for this technical
difference). This is so because typical channel responses like that in (\ref{eq:55})
have spectra that cover the whole frequency axis and their energy is, therefore,
infinite. Additionally, this alternative modeling of $\FH(f)$ requires to measure its
size using the supremum norm and not the energy. So, we require a bound $A$ such that
$|\FH(f)|\leq A$ for any $f$. This bound will have a theoretical use only, given that
it will allow us to derive a proper bound on the estimation error.

We proceed to derive the proposed estimator for $\FH(f)$ in the next section using a
sinc series like (\ref{eq:54}). The starting point will be the RMS error formula in
(\ref{eq:52}). However, we will perform the derivation for a generic bounded band-limited
signal $\Fs(x)$ with spectral support $]-1/2,1/2[$, given that the estimator derived will
be usable whenever any signal of this kind (in any application) must be
estimated from its own nonuniform samples. The problem addressed in the next section is
the following.

\begin{eprob}
Consider a bounded band-limited signal $\Fs(x)$, $|\Fs(x)|\leq A$, with spectral support
lying in {$]-1/2,1/2[$}. Also let $z_m$ denote $M$ noisy samples following the model
\be{eq:47}
z_m=\Fs(x_m)+\Feps_m,
\ee
where the $\Feps_m$ are independent complex Gaussian samples of equal variance
$\sigma_\epsilon^2$ and zero mean, and the abscissas $x_m$ are distinct. The objective is
to estimate $\Fs(x)$ using a linear estimator with coefficients $\Fc_m(x)$, with small
error specified by  
\be{eq:48}
\Big(\FE\big\{\big|\Fs(x)-\summ z_m\Fc_m(x)\big|^2\big\}\Big)^2.
\ee
We view $\Fs(x)$ as deterministic. 
\end{eprob}

This is the problem we have just discussed if we identify the following functions and
variables:
\begin{align}
\label{eq:49}
\Fs(x)&\;\rightarrow\;\FH\Big(\frac{x}{T_h}\Big)\Fe^{j\pi x}& z_m&\;\rightarrow\;
V_m\Fe^{j\pi x_m}\\
\Feps_m&\;\rightarrow\; E_m\Fe^{j\pi x_m}&
\Fc_m(x)&\;\rightarrow\;\Fe^{j \pi (x-x_m)}\Fg_m(x/T_h) \nonumber
\\
\Fx &\;\rightarrow\; f T_h 
\nonumber
\end{align}

\section{Design of the signal estimator}
\label{sec:dse}

Consider the signal $\Fs(x)$ in the previous estimation problem and its sinc series
\be{eq:0}
\Fs(x)=\sump \Fs(p)\sinc(x-p).
\ee
This series is called Zakai's series in the sampling theory literature, and it holds due
to Theorem 6.21 in \cite{Higgins96}, where we view the samples $\Fs(p)$ as deterministic
and bounded, $|\Fs(p)|\leq A$. Note that the bandwidth of $\Fs(t)$ must be strictly
smaller than 1. Actually, there are signals of bandwidth 1, like $\sin(\pi t)$, for which
(\ref{eq:0}) is false. By substituting (\ref{eq:47}) into (\ref{eq:48}) and using
$\FE\{\epsilon_m\}=0$, we obtain:
\begin{multline}
\label{eq:3}
\FE\{|\Fs(x)-\summ z_m\Fc_m(x)|^2\}\\
=\FE\{|\Fs(x)-\summ (\Fs(x_m)+\Feps_m)\Fc_m(x)|^2\}\\
= |\Fs(x)-\summ \Fs(x_m)\Fc_m(x)|^2+\sigma_\epsilon^2\summ|\Fc_m(x)|^2.
\end{multline}
Next, the first term can be bounded using the sinc series in (\ref{eq:0}), noting that
$|\Fs(p)|\leq A$:
\begin{multline}
\label{eq:4}
|\Fs(x)-\summ \Fs(x_m)\Fc_m(x)|^2
= |\sump \Fs(p)\sinc(x-p) \hfill \\ \hfill
 -\summ \sump \Fs(p)\sinc(x_m-p)\Fc_m(x)|^2\\
=\Big|\sump  \Fs(p) \Big(\sinc(x-p)-\summ \sinc(x_m-p)\Fc_m(x)\Big)\Big|^2\\
\leq A^2\sump \Big|\sinc(x-p)-\summ \sinc(x_m-p)\Fc_m(x)\Big|^2.
\end{multline}

By substituting into (\ref{eq:3}), we obtain 
\begin{multline}
\label{eq:8}
\FE\{|\Fs(x)-\summ z_m\Fc_m(x)|^2\}\leq\\
A^2\sump \Big|\sinc(x-p)-\summ \sinc(x_m-p)\Fc_m(x)\Big|^2
\\ \hfill +\sigma_\epsilon^2 \summ|\Fc_m(x)|^2.
\end{multline}
Since we intend to minimize the right-hand side of this inequality, we may assume that
$\Fc_m(x)$ is real, given that $\sinc(x)$ is a real function whenever $x$ is real. Next, 
we may use the property
\be{eq:12}
\sump \sinc(y-p)\sinc(y'-p)=\sinc(y-y'),
\ee
valid for any $y$ and $y'$, to expand the summation's argument in the second line of
(\ref{eq:8}). After straight-forward manipulations, the right-hand side of (\ref{eq:8})
can be written as a quadratic form. In matrix notation, we obtain
\begin{multline}
\label{eq:15}
\FE\{|\Fs(x)-\summ z_m\Fc_m(x)|^2\}\leq\\
A^2\Big(\vc(x)^T\Big(\mg+ \frac{\sigma^2_\epsilon}{A^2}\mi\Big)\vc(x)
-2\vc(x)^T\vg(x)+1\Big),
\end{multline}
where 
\begin{align}
\label{eq:10}
[\mg]_{m,m'}&\equiv \sinc(x_m-x_{m'}),&
[\vg(x)]_m&\equiv\sinc(x-x_m), \nonumber \\
[\vc(x)]_m&\equiv \Fc_m(x), \nonumber
\end{align}
$1\leq m\leq M,\,1\leq m'\leq M$. The minimum of this form is attained at the argument
\be{eq:9}
\hat\vc(x)\equiv \Big(\mg+ \frac{\sigma^2_\epsilon}{A^2}\mi\Big)^{-1}\vg(x)
\ee
and the corresponding bound in (\ref{eq:15}) is
\begin{multline}
\label{eq:16}
\FE\{|\Fs(x)-\summ z_m\Fc_m(x)|^2\} \\
\leq A^2\Bigg(1-\vg(x)^T
\Big(\mg+ \frac{\sigma^2_\epsilon}{A^2}\mi\Big)^{-1}\vg(x)\Bigg).
\end{multline}

In summary, if we place the samples $z_m$ in a vector $[\vz]_m\equiv z_m$, we have
obtained the following linear estimator for $\Fs(x)$,
\be{eq:28}
\Fs(x)\approx \vg(x)^T\Big(\mg+ \frac{\sigma^2_\epsilon}{A^2}\mi\Big)^{-1}\vz.
\ee
The application of the replacements in (\ref{eq:49}) to this formula yields the desired
estimator for $\FH(f)$. 

\section{Numerical example}
\label{sec:ne}

In order to assess the proposed estimator, we proceed to compare it with the deterministic
ML estimator based on the TDL model. For implementing this last estimator, we have used
the model in \cite[Sec. III.A]{Morelli01}. We consider the following scenario.

\paragraph{OFDM signal} We employ the OFDM signal in \cite[Sec. IV.C]{Morelli01} with the
  following parameters,
\begin{itemize}
\item DFT size: 512.
\item Number of modulated carriers:  433.
\item Number of pilots: $M=28$.
\item Indices of pilot carriers: $i_m=16 m$, $m=0,$ $1,$ $\ldots,$ $M-1$.
\item For simplicity, we take the frequency spacing $\Delta f=1$.
\end{itemize}

\paragraph{Channel model} 
\label{sec:cm}

The usual numerical examples for assessing the RMS error in the
literature describe the channel's response using a tap delay line, whose weights have a
specific distribution; (see for example \cite{Morelli01}). However, the estimator proposed
in this letter has been designed for a specific maximum delay spread $T_h$, and this
parameter can be hardly obtained from a TDL model. Therefore, to assess the proposed
estimator we have generated channel impulse responses with a given maximum time
spread $T_h$. Specifically, we have used channel impulse responses of the form in
(\ref{eq:55}), where
\begin{itemize}
\item $K-1$ has a Poisson distribution of parameter $\lambda=9$.
\item $a_k$ are independent complex Gaussian variables of zero mean and variance
  $\sigma_a^2\Fe^{-2(k-1)/K}$.
\item The delays $\tau_k$ are uniformly distributed in $[0,T_h]$.

\item $\sigma_a^2$ is selected numerically so that
\be{eq:36}
\FE\{|\FH(f)|^2\}=1
\ee
for any $f$.

\end{itemize}

\paragraph{Signal-to-noise ratio} We define the SNR as 
\be{eq:33}
\gamma\equiv \frac{\FE\{|\FH(f)|^2\}}{\sigma_E^2},
\ee
where $\sigma_E^2$ is the variance of $E_n$ in (\ref{eq:50}). Note that
(\ref{eq:36}) implies $\sigma_E^2=1/\gamma$. We set $\gamma=30$ dB.

\medskip
\noindent
\paragraph{Delay spread} We select the channel duration $T_h$ as a function of the pilot
  average spacing. Specifically, if
\be{eq:35} 
B_{av}\equiv \frac{i_{M-1}-i_0}{M-1}\Delta f,
\ee
then we set $T_h=\alpha/B_{av}$ for $0\leq\alpha\leq 1$.

\begin{figure}
\subfigure[$T_h=0.25/B_{av}$]{\label{fig:2}\includegraphics{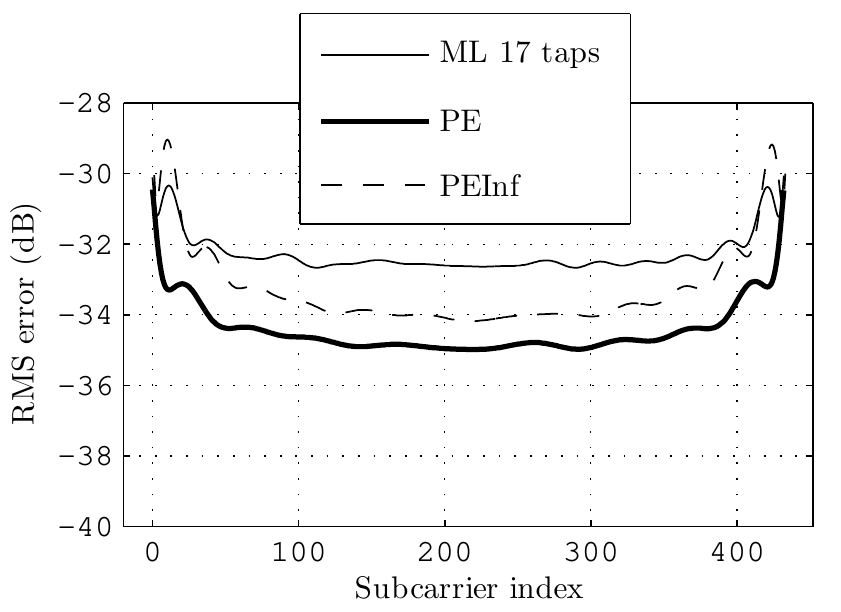}}
\subfigure[$T_h=0.125/B_{av}$]{\label{fig:3}\includegraphics{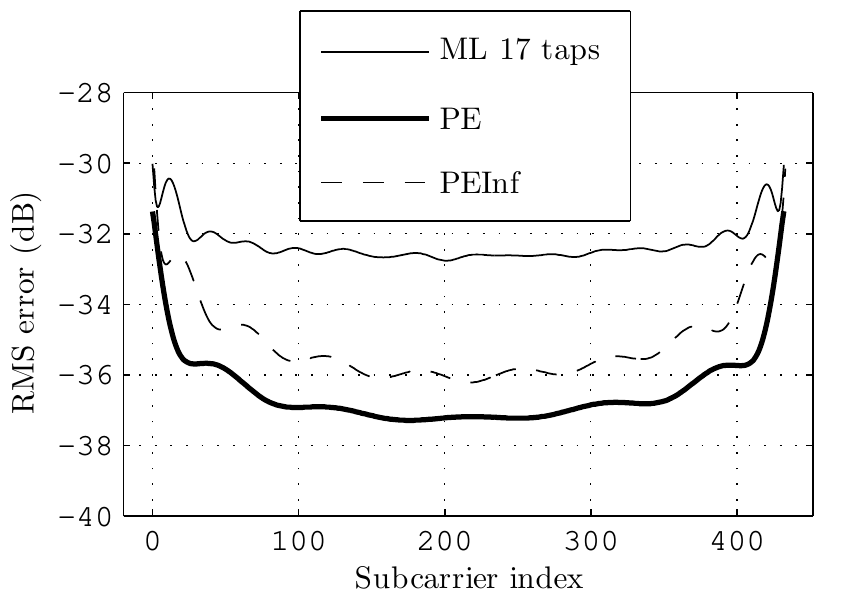}}
\caption{RMS error for the ML, PE, and PEInf estimators.}
\end{figure}

\smallskip
Figs. \ref{fig:2} and \ref{fig:3} show the RMS error for $T_h=0.25/B_{av}$ and
$T_h=0.125/B_{av}$, and for the following three estimators: 

\begin{itemize}
\item ML: ML estimator with the number of taps providing the smallest RMS error.
\item PE: Proposed estimator in (\ref{eq:28})  with the replacements in (\ref{eq:49}).
\item PEInf: This last estimator but setting $\gamma=\infty$.
\end{itemize}

We can see that estimator PE performs significantly better than estimator ML, and that the
improvement is larger for $\alpha=0.125$. In average the improvement is $2.6$ dB and $4.2$
dB for $T_h=0.25/B_{av}$ and $T_h=0.125/B_{av}$, respectively. Estimator PEInf also
outperforms estimator ML, though with a somewhat larger RMS error.

\begin{figure}
\includegraphics[scale=0.7]{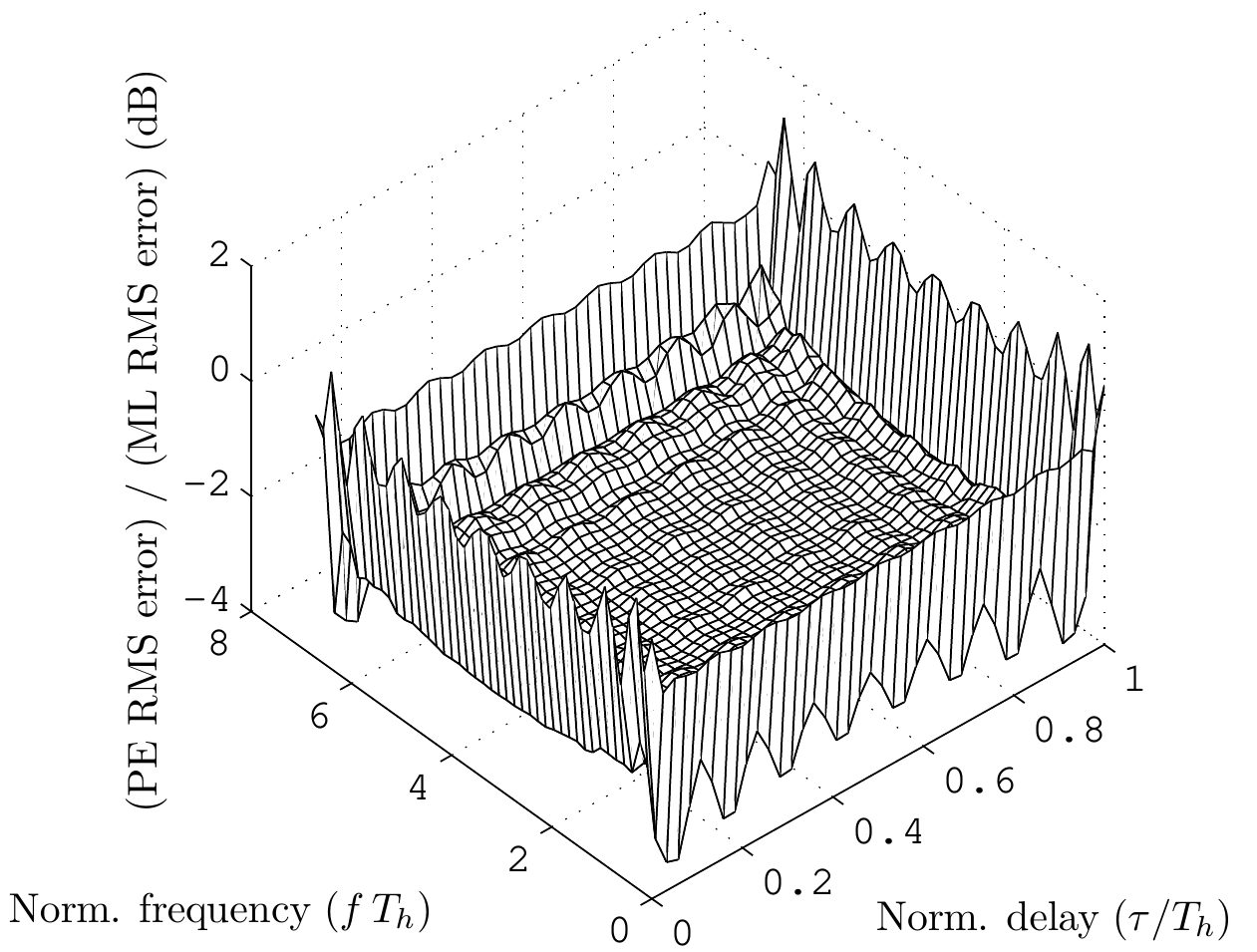}
\caption{\label{fig:6} RMS error of proposed method relative to RMS error of TDL estimator
  for all possible delays and frequencies.}
\end{figure}

Fig. \ref{fig:6} shows the reduction in RMS error of the estimator PE relative to
the estimator ML. For each possible delay $\tau$, we can see in this figure the
reduction in RMS error brought by estimator PE at frequency $f$ when the
channel is $\Fh(t)= \delta (t-\tau)$. Except at the limit delays and frequencies, the
reduction in RMS error is 2.4 dB roughly.


\section{Conclusions}

We have recalled the problem of estimating a channel spectrum from a finite number of
nonuniform samples. This problem appears in OFDM system equipped with pilot carriers,
after the usual DFT processing. We have shown that this problem can be cast as that of
estimating a band-limited signal from nonuniform samples through a proper normalization.
Afterward, we have derived an estimator in which the performance measure is the RMS error,
assuming that the signal (or channel spectrum) is bounded and the samples are contaminated
by independent zero-mean complex Gaussian noise samples of equal variance. Finally, we
have compared this estimator with the usual ML estimator based on a TDL model, in order to
assess its performance in a basic OFDM setting. The main conclusion is that the proposed
estimator improves on the ML estimator in RMS error significantly, provided there is some
spectral oversampling.

\bibliographystyle{IEEEbib}

\bibliography{../../../Utilities/LaTeX/Bibliography}

\end{document}